# Impact of WACC on Firm Profitability: Evidence from Food and Allied Industry of Bangladesh

Farhana Rahman

## ABSTRACT

**The research paper aims to analyze the underlying relationship in between the profitability and cost of funds of a firm. A total of twelve companies were selected as a sample for this study which are listed in Dhaka Stock Exchange under Food and Allied Industry. A panel data set of 15 years from 2005 to 2019 was used to conduct the necessary analysis. In this paper, Return on Asset (ROA) is used as the accounting criteria of profitability. WACC is the independent variable while Firm Size, Firm Age and Firm Leverage are used as control variable for the study. Fixed Effects Panel Regression Model is used to analyze the dataset. The result of the analysis shows that WACC is negatively related with the profitability measure and this relationship is significant. The study has potential to be replicated by other industries like textile, cement, pharmaceutical & chemical, fuel & power, tannery etc.**

**Keywords:** Financial Leverage, Return on Asset (ROA), Weighted Average Cost of Capital (WACC).

Author
Assistant Professor, Department of Organization Strategy & Leadership, University of Dhaka, Bangladesh. (farhanarahman@du.ac.bd)

## I. INTRODUCTION

The analysis on the cost of capital and its relationship with the profitability of a company receives an extensive level of focus in the financial management (Gitman, 2006). A careful review of the firm's cost of capital and its underlying relationship with the profitability of the company is crucial in the investment decision.

A point where the cost of capital of the company is minimum and return from the investment projects are maximum, optimal capital structure is recognized. The profitability of the company is highest at this optimal capital structure point.

When assessing an investment project, a firm may evaluate a project's return in terms of its cost of finance. When the return of the project will exceed the company's cost of finance, acceptance of the project will improve the overall profitability of the company and increase the value of the firm. On the other hand, when the return of the project will not exceed the company's cost of finance, acceptance of the project will deteriorate the overall profitability of the company and squeeze the value.

The cost of capital and its underlying relationship framework can be used to assess the financial performance of the top management. This kind of evaluation entails a comparison of the profitability of the investment project (undertaken by the top management of the firm) with the overall cost of capital. A positive Net Present Value yielding project makes a net contribution to the shareholder's wealth.

The cost of funds (WACC) is an important determinant of the profitability. The investment decision of the company highly depends upon the cost of the fund. So, it is very necessary to understand how cost of funds (WACC) affects profitability of the company.

## II. RESEARCH OBJECTIVES

The key objective of this study is as follows:
- To understand underlying relationship in between the profitability measure Return on Asset (ROA) and the overall cost of funds (WACC) of the firm.
- To understand how the cost of funds affects (positively or negatively) the profitability of the firm.

## III. LITERATURE REVIEW

Capital structure is a way to finance the assets of a company through some combination of equity, debt, or hybrid securities. A firm's capital structure is the composition of its liabilities (Khadka, 2007). Capital structure theory, also popularly known as the leverage theory, concentrates on the capital structure components and determination of an optimal capital structure for a firm (Gitman, 1994). They help to understand how the mix of equity and debt will have an impact on the value of the organization. A number of capital structure theories have talked about the optimal mix of debt and equity. According to these theories, there are costs and benefits associated with debt as well as equity. A company should always choose the value maximizing combination of debt and equity (Welch, 2005).

The incremental benefits of raising debt in the capital structure decreases as the level of debt increases (Ross, 2007). Using equity capital may incur more cost compared to the debt as it is ineligible for the tax savings advantage, but at the higher level of debt it may be more attractive as it does not bring out any financial risk like as debt capital. So, a firm which is looking for enhancing its overall value should focus on trade off during the time of making financing decision.

According to Khadka (2007), the cost of capital refers to what a firm should pay for the capital used to finance new investments. Theoretically, there is a negative relationship between the profitability of the firm and its WACC. WACC is served as the discount rate of the projects undertaken by the firm (Ross, 2007). Usually higher discount rate results in less cash flows and this ultimately implies low NPV project. Low NPV project results in the declining of the profitability of the firm (Miglo, 2012). Several capital structure theories and empirical studies have tried to explore the relationship between the profitability of the firm and WACC.

According to theory of Modigliani-Miller, firm value is determined through profitability, irrespective of the capital structure. There is no direct link in between the profitability and the debt structure as well as the WACC of the firm (Higgins, 2005).

Modigliani and Miller (1963) later modified their original MM's model and considered the tax deductibility of interest (tax shields effect) thus demonstrate that the market value of a firm is an increasing function of leverage with the existence of corporate tax that allow the deductibility of interest payments. Later supported by Brigham and Gapenski (1996) which argue that an optimal capital structure can be attained if there exist a tax sheltering benefit, provided an increase in debt level is equal to the bankruptcy costs.

It is very much easy to obtain an optimal capital structure if there remains an arrangement of a tax sheltering benefits. Brigham and Gapenski (1996) expressed their views supporting this phenomenon. Brigham and Gapenski (1996) also asserted that, it may become easier if it can be ensured that, an increase in the level of organizational debt is equal to the costs that are related with bankruptcy. It also should be taken under consideration that, managers of the organization have the abilities of to identify the proper time for optimal capital structure and also have the abilities to maintain this at the expected level. In this point, the cost of financing and capital can be lessened. On the other hand, the profit potentials can be increased in this regard. From this perspective, it can be said that, there exists a negative relationship between weighted average cost of capital and profit potentialities of an organization.

According to the study conducted by Tashfeen and Liton (2010), there exist a strong negative correlation between the cost of capital of commercial banks and their respective returns. The analysis was conducted on the 24 listed commercial banks in the Dhaka Stock Exchange, Bangladesh between January, 2006 and December, 2008.

Hussain, Ali & Islam (2012) found a negative association between return on equity and weighed average cost of capital. Their research has applied the weighted average cost of capital vis-a-vis risk premium model, Gordon model, as well as FAMA & French Model to the Cement Industry from Pakistan. The outcomes have measured the balanced impact of cost of capital on return on equity in Cement industry under the assumption that level of managerial as well as operation competency in all the firms working in this industry continues to be unchanged.

Theoretically, there is an inverse relationship between WACC and the profitability of the firm. Higher (lower) WACC results in higher (lower) discount rate for the firm's project & lower (higher) profitability (Higgins, 2005). The past empirical studies conducted to explore this relationship have also agreed with the theoretical claim.

## IV. RESEARCH METHODOLOGY

In this study, the causal relationship between the ROA and WACC has been assessed to make a conclusion whether there is a relationship between the weighted average cost of capital and profitability of a company. The control variables in this study are firm size, firm age and firm leverage. The study is largely quantitative in such a way that financial data have been collected for the period from 2005 to 2019 for the 12 respective companies belonging to the food and allied industry of Bangladesh.

The statistical models for assessing the underlying relationship in between the cost of funds of the firm (WACC) and profitability measure of the company is as follows.

$$ROA_{it} = \beta_0 + \beta_1 WACC_{it} + \beta_2 FIRM\_SIZE_{it} + \beta_3 FIRM\_AGE_{it} + \beta_4 FIRM\_LEV_{it} + \varepsilon_{it} \quad (1)$$

In this model,

$ROA_{it}$ = Return on asset of firm i for period t

$WACC_{it}$ = Weighted average cost of capital of the firm i for period t

$FIRM\_SIZE_{it}$ = Firm Size of firm i for period t

$FIRM\_AGE_{it}$ = Age of firm i for period t

$FIRM\_LEV_{it}$ = Financial leverage of firm i for period t

The null and alternate hypothesis for the model is-

$H_0$: There is no significant relationship between ROA and WACC.

$H_1$: There is significant relationship between ROA and WACC.

## V. DATA DESCRIPTION

This paper develops estimates of the ROA and weighted cost of capital (WACC) for 12 companies of the Bangladesh food and allied industry. These estimations will be based on the financial data from 2005 to 2019. These 12 retail companies are selected random basis. Random sampling refers to a sampling technique where each item of the entire set population has equal probability to be selected. The variables are-

**Return on Asset (ROA):** Return on Asset (ROA) is the dependent variable of the study. ROA is a profitability measure that shows how profitable a company is relative to its total asset. ROA is calculated with the following formula-

Profit after tax/Total Asset ( Ibrahim, Abdulkarim, Muktar, Gurama & Peter, 2021)

**Weighted Average Cost of Capital (WACC):** Corporation's weighted average cost of capital is the weighted average of cost each individual capital source namely the common equity, preferred stock, and debt. Weighted average cost of capital (WACC) is calculated in the following way-

$$WACC = [W_E \times C_E] + [W_D \times (1-t) \times C_D] + [W_P \times C_P]$$

Here,

$C_E$: Cost of Equity
$C_D$: Cost of debt
$C_P$: Cost of Preferred Stock
$W_E$: Share of common equity in capital employed
$W_D$: Share of debt in capital employed
$W_P$: Share of preferred stock in capital employed
t : Effective tax rate.

**Firm Size:** Firm size is measured by the natural log of total asset value. (Alrjoub & Ahmad, 2017) It is expected that larger firm can attain debt at a lower cost due to its creditworthiness and lower likelihood of bankruptcy. The variable is expressed in natural logarithm form to normalize the data for analysis purpose. (Rahman, 2017)

**Firm Age**: Firm age is considered as the date of listing to the year of observation (Ajay & Madhumathi, 2012; Ibrahim et all, 2021).

**Firm Leverage**: Firm leverage is calculated by total debt divided by total asset (Alrjoub & Ahmad, 2017).

## VI. EMPIRICAL RESULTS

### A. Descriptive Statistics

Table I represents the mean, standard deviation, minimum values and maximum values of the variables used in the study. A balanced panel data is used for the study. The mean value for ROA of listed companies in the food and allied industries of Bangladesh is 0.147 indicating that those firms earn an average return of 14.7% on their asset. The minimum and maximum values of the dependent variable are of -0. .467 and 0.844 respectively. On the other hand, the weighted average cost of capital is 11.8% on average for the study period. The control variable Firm age has a mean of 18.678 years while the maximum firm age in our sample is 32 years. 0.645 mean of Firm Leverage variable indicates that on an average the firms employed approximately 64.5% debt to finance the total asset of the firm.

TABLE I: DESCRIPTIVE STATISTICS

| Variable | Mean | Standard Deviation | Min | Max |
|---|---|---|---|---|
| ROA | 0.147 | 1.343 | -.467 | 0.844 |
| WACC | 0.118 | 0.458 | -0.218 | 0.675 |
| FIRM AGE | 18.678 | 6.247 | 9 | 32 |
| FIRM SIZE | 12.675 | 0.896 | 3.473 | 14.764 |
| FIRM LEV | 0.645 | 0.387 | .183 | 1.36 |

### B. Correlation

Table II represents the correlation coefficients among the variables of interest. However, it is found that none of the variables have strong association. The correlation coefficient of ROA and WACC is found to be -0.26. Firm size and firm leverage has a correlation of -0.23 indicating a weak negative correlation between the variables. Otherwise the correlation coefficient among other variables are minimal.

TABLE II: CORRELATION MATRIX

| | ROA | WACC | FIRM AGE | FIRM SIZE | FIRM LEVERAGE |
|---|---|---|---|---|---|
| ROA | 1.00 | | | | |
| WACC | -0.26 | 1.00 | | | |
| FIRM AGE | -0.11 | -0.09 | 1.00 | | |
| FIRM SIZE | -0.18 | -0.04 | 0.04 | 1.00 | |
| FIRM LEVERAGE | 0.09 | -0.05 | 0.08 | -0.23 | 1.00 |

### C. Regression Output

Panel data methodology has been used to check the impact of WACC on the firm profitability. The underlying assumption behind using panel data is that there is heterogeneity among the firms. Also, panel data set is more informative and efficient. Hausman test is used to identify whether fixed effects model or random effects model is appropriate (Rahman, 2022). Table III represents the test results of Hausman test. The p value of Hausman test results indicates to the rejection of null hypothesis which is random effects model is appropriate. So, fixed effects model is chosen.

TABLE III: HAUSMAN TEST

| Hausman (1978) specification test | Chi-Sq. Statistic | Prob. |
|---|---|---|
| | 34.469 | 0.0000 |

TABLE IV: FIXED EFFECTS MODEL OUTPUT

| | Coefficient | St. Err. | p-value |
|---|---|---|---|
| WACC | -0.097 | 0.039 | .002*** |
| FIRM AGE | -0.002 | 1.463 | .358 |
| FIRM SIZE | -0.139 | 0.251 | .126 |
| FIRM LEV | 0.035 | 0.025 | .067* |
| Constant | .053 | 1.361 | .165 |
| Within R-squared | 0.360 | F-test | 12.374 |
| Number of obs | 180 | Prob > F | 0.000 |
| *** $p<.01$, ** $p<.05$, * $p<.1$ | | | |

Table IV reports the regression output. WACC is found to have significant negative impact on the firm profitability at 1% level of significance. In other words, as WACC increases, the ROA decreases. The result is consistent with literature. Firm leverage is also found to have a significant impact on profitability. However, the other control variables are found to have insignificant impact.

## VII. CONCLUSION

In this study we have done a research analysis to explore the underlying relationship in between the profitability and WACC of a company. The study has employed a model specification for the purpose of testing the postulate hypothesis, where Return on Asset (ROA) is used as the profitability measure along with cost of capital measure WACC as the independent variable. Firm Size, Firm Age and Firm leverage are used as control variables in the study.

The regression result shows that there is a negative relationship between the Return on Asset (ROA) and WACC thus null hypothesis of 'There is no significant relationship in between the ROA and WACC' is refuted. This result is consistent with the findings of Nadya et al. (2019), Zheng et al. (2017) and Ibrahim et al. (2021) who found that firm's cost of capital has a negative and significant impact on firm financial performance. Higher (lower) WACC results in higher (lower) discount rate for the firm's project & lower (higher) profitability The negative relationship might also be due to the fact that, if the equity is kept constant (considering equity issue is a rare phenomenon in Bangladesh), as more debt is issued, the financial distress of the firm increases. Employing extra debt than optimum increases the cost of financial distress more than the benefit received from interest tax shield. Thus, increase in the weighted average cost of capital reduces the firm profitability.

Our findings add to the existing literature regarding the effect of cost of capital on the financial performance of the firm which is a valuable information to the finance providers as it helps them to make their investment decisions prudently. The result of this analysis can be used as a reference in future researches related to similar topics. The consideration of other factors of profitability along with the WACC will provide a scope to understand the relative impact of the WACC on the profitability. There is also a scope to replicate this research for other industries of Bangladesh like textile, cement, pharmaceutical & chemicals, fuel & power, tannery etc.